\documentclass[fleqn,twoside]{article}
\usepackage{espcrc2}

\usepackage{graphicx}
\usepackage[figuresright]{rotating}

\newcommand{\AmS}{{\protect\the\textfont2
  A\kern-.1667em\lower.5ex\hbox{M}\kern-.125emS}}

\let\ga=\gamma

\let\Om=\Omega

\def\to{\rightarrow}

\let\<=\langle
\let\>=\rangle

\let\ad=\dagger

\def\beq{\begin{equation}}
\def\eeq{\end{equation}}
\def\ba{\begin{array}}
\def\bea{\begin{eqnarray}}
\def\ea{\end{array}}
\def\eea{\end{eqnarray}}
\def\bit{\begin{itemize}}
\def\eit{\end{itemize}}

\def\comment#1{ \hbox{[{\it Comment suppressed here.}\/]} }
\def\hide#1{}
      
\def\O{ {\cal O} }

\def\Tr{\hbox{Tr}}

\def\pbf{{\bf p}}

\def\xbf{{\bf x}}
\def\ybf{{\bf y}}
\def\zerobf{{\bf 0}}

\def\={\!=\!}
\def\+{\,+\,}
\def\-{\,-\,}

\title{Semileptonic decays of $D$ mesons in unquenched lattice QCD\thanks{
Talk presented by M.~Okamoto.}}

\author{Masataka~Okamoto\address[FNAL]{
Fermi National Accelerator Laboratory, P.O. Box 500, Batavia, IL 60510},
Massimo~di~Pierro\address[DEPAUL]{
School of Computer Science, Telecommunications and Information
Systems, DePaul University, Chicago, Illinois 60604},
Aida~X.~El-Khadra\address[UIUC]{
Department of Physics, University of Illinois, Urbana, IL 61801},
Steven~Gottlieb\address[IU]{
Department of Physics, Indiana University, Bloomington, IN 47405},
Andreas~S.~Kronfeld\addressmark[FNAL], 
Paul~B.~Mackenzie\addressmark[FNAL],
Damian~P.~Menscher\addressmark[UIUC], Mehmet~B.~Oktay\addressmark[UIUC],
and James~N.~Simone\addressmark[FNAL]
}

\begin{document}

\begin{abstract}
We present our preliminary results for semileptonic
form factors of $D$ mesons in unquenched lattice QCD.
Simulations are carried out with $n_f=2+1$ dynamical quarks
using gauge configurations generated by the MILC collaboration.
For the valence quarks, we adopt an improved staggered light
quark action and the clover heavy quark action.
Our results for $D \to K$ and $D \to \pi$ form factors at $q^2=0$
are in agreement with the experimental values.
\end{abstract}

\maketitle

\section{INTRODUCTION}

In order to extract the CKM matrix element $|V_{ub}|$ accurately
with experimental measurement of the semileptonic decay width, 
a precision lattice QCD calculation of the $B \to \pi$ form factor
is required. 
To check the reliability of lattice calculations of the heavy to 
light form factors, study of the semileptonic decays of $D$ mesons,
such as $D \to K$ and $D \to \pi$, 
is a good test ground because the corresponding CKM matrices 
$|V_{cs}|$ and $|V_{cd}|$ are known more accurately than $|V_{ub}|$.
Furthermore, forthcoming experiments by the CLEO-c collaboration 
will provide more stringent checks of lattice calculations in
the $D$ meson system.

We have started new lattice calculations of heavy quark physics
in unquenched QCD\cite{spectrum}, and here we report our 
preliminary results for semileptonic form factors of $D$ mesons.
We are using unquenched gauge configurations with $n_f=2+1$ 
improved staggered quarks generated by the MILC 
collaboration\cite{milc}, with which the systematic errors due to 
the quenched approximation should be almost absent. 
For the valence light quarks, we adopt an improved staggered 
quark action, which allow us to simulate at lighter quark mass than
previous studies with the Wilson-type light quarks.
Hence, our new calculations should have a better control over the chiral
extrapolations.

\section{METHOD}\label{sec:method}

In order to combine the staggered 
light quark with the Wilson-type heavy quark in heavy-light bilinears,
we convert the staggered quark propagator $g(x,y)$
to the ``naive'' quark propagator $G(x,y)$ according to
\beq
g(x,y) \ \Om(x)\ \Om^\ad(y) 
~=~ G(x,y)
\eeq
with 
$
\Om(x) ~=~ \ga_0^{x_0} \ga_1^{x_1} \ga_2^{x_2} \ga_3^{x_3}
$\cite{Wingate:2002fh}.
The 3-point function for the matrix element is then 
computed as 
\bea
C_{3,\mu}^{D\to\pi}(t_x,t_y; \pbf_\pi,\pbf_D) 
= \sum_{\xbf,\ybf} e^{i(\pbf_D-\pbf_\pi)\ybf - i\pbf_D\xbf} \times \nonumber \\
\< \Tr [g_d^\ad(y,0)\Om^\ad(y)\ga_5\ga_\mu G_c(y,x)\ga_5\Om(x)g_u(x,0) ] \> ,
\nonumber 
\eea
where subscripts $d,c,u$ denote quark flavors. 
The matrix element can be extracted from the ratio
\beq
\< \pi|V_\mu |D \> \stackrel{t_x\gg t_y\gg 0}{\sim}
\frac{C_{3,\mu}^{D\to\pi}(t_x,t_y; \pbf_\pi,\pbf_D)}
{C_2^{\pi}(t_y, \pbf_\pi) ~C_2^{D}(t_x-t_y, \pbf_D)}
\eeq
with the $D$ meson (Wilson-Naive) 2-point function $C_2^{D}$
and the pion (Naive-Naive) 2-point function $C_2^{\pi}$.
Care is needed, however, for the overall normalization of amplitude
since the naive quark action describes 16 fermions, which can
cause the doubling of these correlation functions. 

In Ref.~\cite{Wingate:2002fh} it is shown that 
the Wilson-Naive 2-point function $C_2^{D}$ does not have the doubling 
because contributions of quarks with momentum $p\sim\O(\pi /a)$
are suppressed by the Wilson term. The same also holds for the 
3-point functions which include at least one Wilson propagator
such as $C_{3,\mu}^{D\to\pi}$.
On the other hand, the Naive-Naive 2-point function $C_2^{\pi}$
should have 16 equivalent contributions.
Therefore one has to divide it by 16 to get the physical amplitude;
$C_2^{\pi, \rm phys} = C_2^{\pi} / 16$.

\begin{table}[t]
\caption{Quark mass, statistics and the sink time.}
\begin{tabular}{llll}
\hline\hline
$m_l^{\rm sea}/m_s^{\rm sea}$ &$m_l^{\rm val}/m_s^{\rm val}$ & conf & $t_x$ \\
\hline
0.01/0.05 & 0.01/0.0415 & $552\times 4$ & 20\\ 
0.02/0.05 & 0.02/0.0415 & 460 & 20 \\ 
0.03/0.05 & 0.03/0.0415 & 358 & 22 \\ 
0.01/0.05 & 0.0415/0.0415 & 412 & 26   \\ 
$\infty /\infty$ & 0.0415/0.0415 & 350 & 16   \\ 
\hline\hline
\end{tabular}
\vspace{-.6cm}
\label{tab:params}
\end{table}

\section{SIMULATION}

Unquenched calculations are performed using $n_f=2+1$ dynamical 
gauge configurations obtained with an improved staggered quark 
action on a $20^3\times64$ 
lattice ($a^{-1}\!\!\approx\!\! 1.58$ GeV)\cite{milc}. 
For the valence light quarks we use the same staggered 
quark action as for the dynamical quarks.
The valence light quark ($u,d$) mass $m_l^{\rm val}$ is usually set 
equal to the dynamical light quark mass $m_l^{\rm sea}$.
For the valence charm quark we use the clover action 
with the Fermilab interpretation\cite{kkm}.
The hopping parameter is fixed to $K_{\rm charm}=0.119$, based on
our spectrum study\cite{spectrum}.
The $O(a)$ rotation\cite{kkm} is performed for the vector current.

The 3-point functions are computed in the $D$ meson rest frame 
($\pbf_D=\zerobf$) for the light meson momentum $\pbf_\pi$ up to $(1,1,1)$
in lattice units, using local source and sink.
The sink time is fixed to $t_x=20-26$ depending on $m_l^{\rm val}$,
whereas the source time is set to $t_0=0$ with an exception at 
$m_l^{\rm val}=0.01$, where we average over results from four 
source times $t_0=0,16,32$ and 48.
Some simulation parameters are summarized in Table~\ref{tab:params}.

In addition to the unquenched calculations, we also perform 
a quenched simulation at $m_l^{\rm val}=m_s^{\rm val}=0.0415$
using $\beta=5.9$ ($a^{-1}\!\! \approx \!\! 1.80$ GeV) configurations 
on a $16^3\times 32$ lattice used in our previous 
study\cite{El-Khadra:2001rv}.
Comparison between the quenched result with the staggered light quarks
and that with the Wilson-type light quarks\cite{El-Khadra:2001rv}
allows us to check the validity of 
our new calculations.

For the vector current renormalization $Z_{V_\mu}^{cd}$
we follow the method in Ref.~\cite{El-Khadra:2001rv}.
We take 
$
Z_{V_\mu}^{cd}\!\!=\!\!\rho_{V_\mu}(Z_{V}^{cc} Z_{V}^{dd})^{1/2} ,
$
where $Z_{V}^{qq}$ ($q\!\!=\!\!c,d$) is the renormalization constant for 
the flavor-conserving current, which we compute nonperturbatively
from the charge normalization condition
$Z_{V}^{qq} \< D(\zerobf)| V_4^{qq} | D(\zerobf)\>\!\! =\!\! 2 m_D$.
The $\rho_{V_\mu}$
is set to unity.
The one-loop calculation is in progress.

\begin{figure}[t]
\includegraphics*[width=7.5cm]{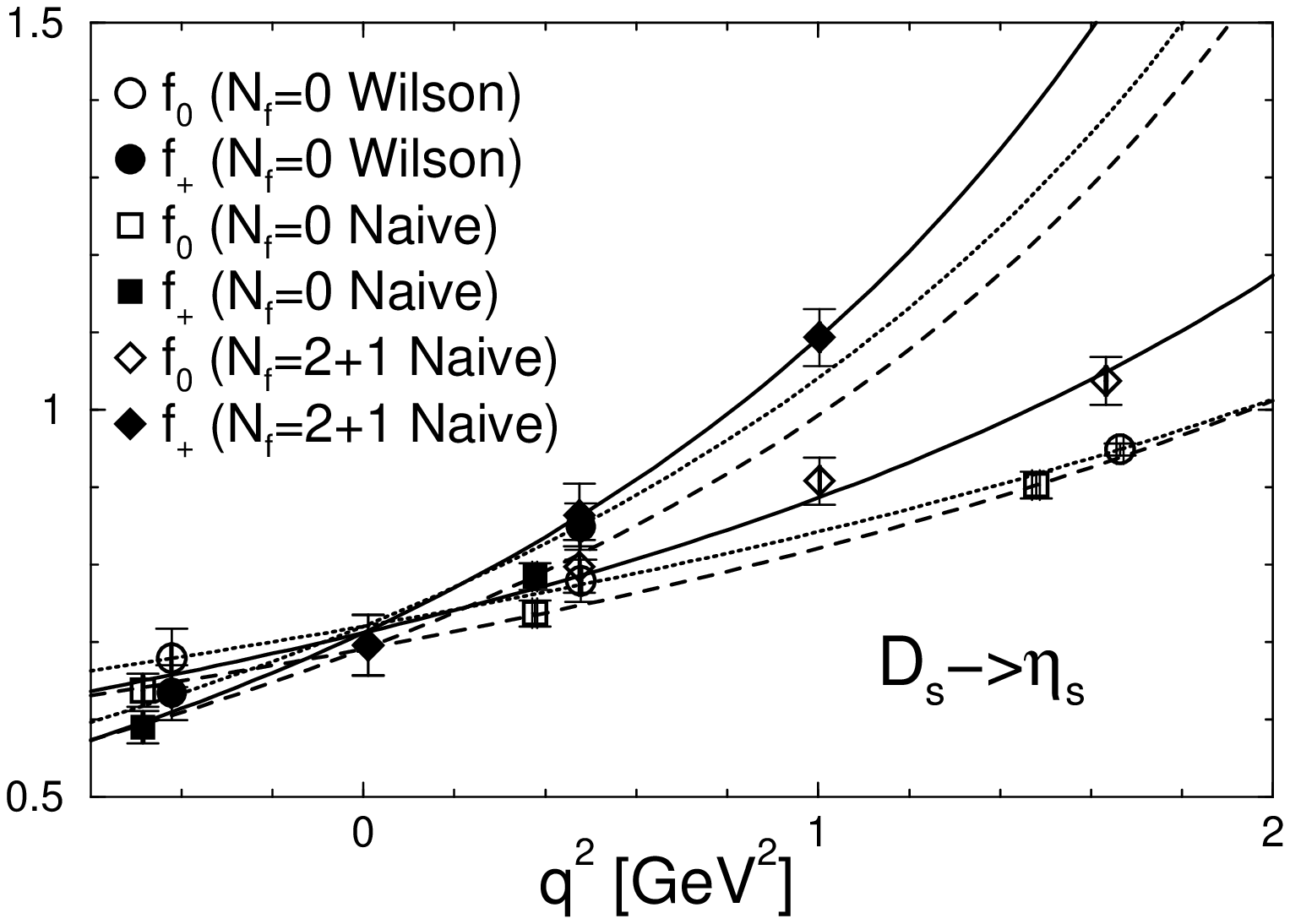}
\vspace{-1.5cm}
\caption{$D_s\to\eta_s(s\bar{s})$ form factors.}
\vspace{-.72cm}
\label{fig:Ds2eta}
\end{figure}

\section{RESULTS}

Form factors are defined through
\bea
\< \pi | V^\mu | D \>
\!\!\!\! &=& \!\!\!\!
f_+(q^2) \left[p_D+p_\pi-\frac{m_D^2-m_\pi^2}{q^2}\, q\right]^\mu \nonumber\\
&& \!\!\!\! ~+~ f_0(q^2) \, \frac{m_D^2-m_\pi^2}{q^2} \, q^\mu \nonumber\\
&=& \!\!\!\!
\sqrt{2m_D} \, \left[v^\mu \, f_\parallel(E) +
p^\mu_\perp \, f_\perp(E) \right] \nonumber
\eea
with $q = p_D - p_\pi$, $v=p_D/m_D$, $p_\perp=p_\pi-Ev$ and $E=E_\pi$.
The second expression using $f_\parallel$ and $f_\perp$ is more convenient
when one considers the heavy quark expansion and the chiral limit.

\subsection{$D_s\to\eta_s$}

In Fig.~\ref{fig:Ds2eta} we summarize the results of form factors 
$f_0$ and $f_+$ for the $D_s\to\eta_s(s\bar{s})$ decay obtained with 
the naive(staggered) light quarks and previous results\cite{El-Khadra:2001rv}
with the Wilson-type light quarks. 
The lines in the figure are fitting curves with a parametrization by 
Becirevic and Kaidalov (BK)\cite{Becirevic:1999kt}.
One can see a nice agreement between the quenched result with
the naive quarks (squares, dashed line) 
and that with the Wilson-type quarks (circles, dotted), showing that 
our new method works well.
We also note that the unquenched result (diamonds, solid)
is larger than quenched ones for $q^2 > 0$.
See also Ref.~\cite{detar03} for a similar comparison for the 
$B_s\to\eta_s$ form factors.

\begin{figure}[t]
\centerline{
\includegraphics*[width=3.6cm]{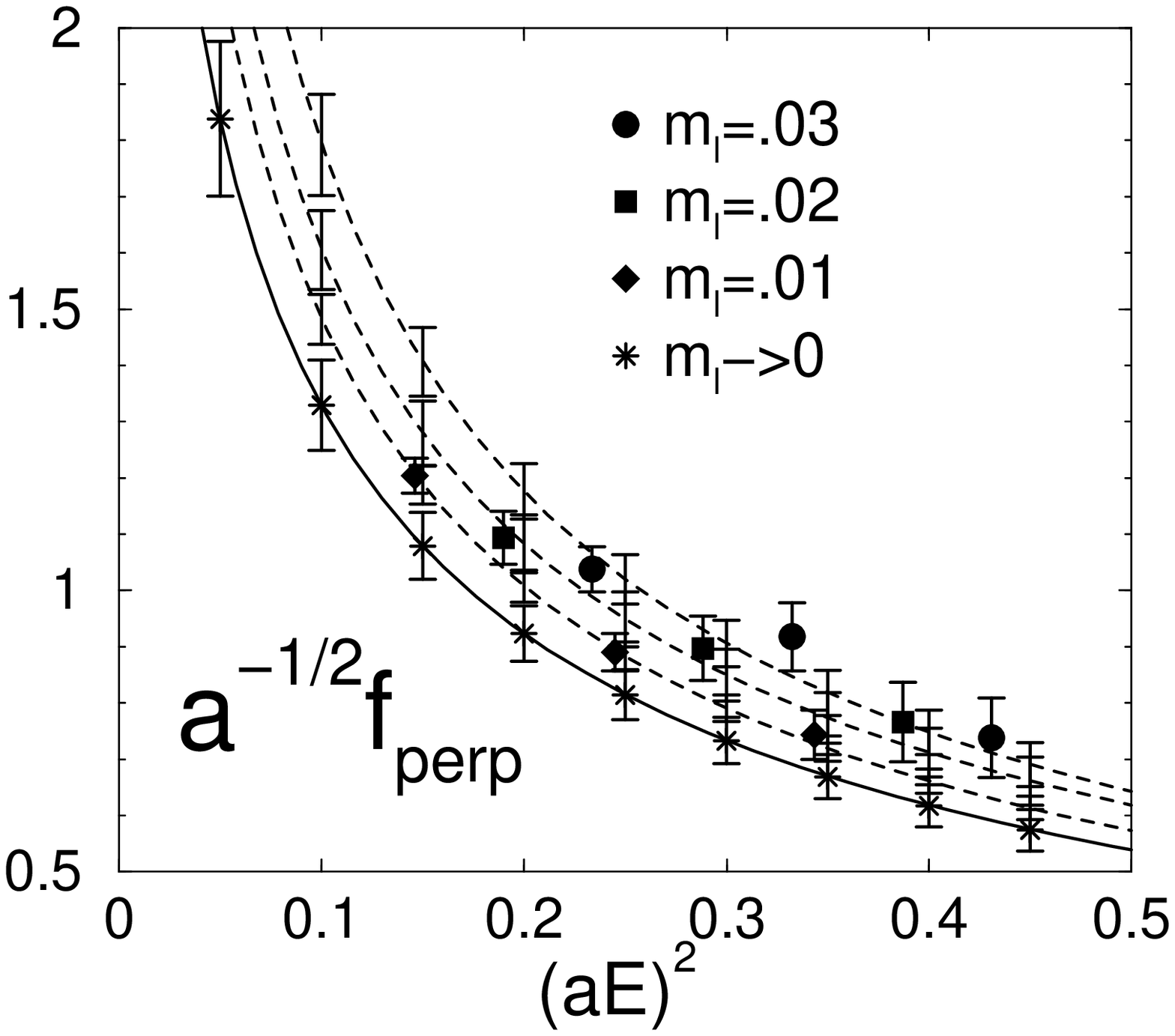}
\includegraphics*[width=3.6cm]{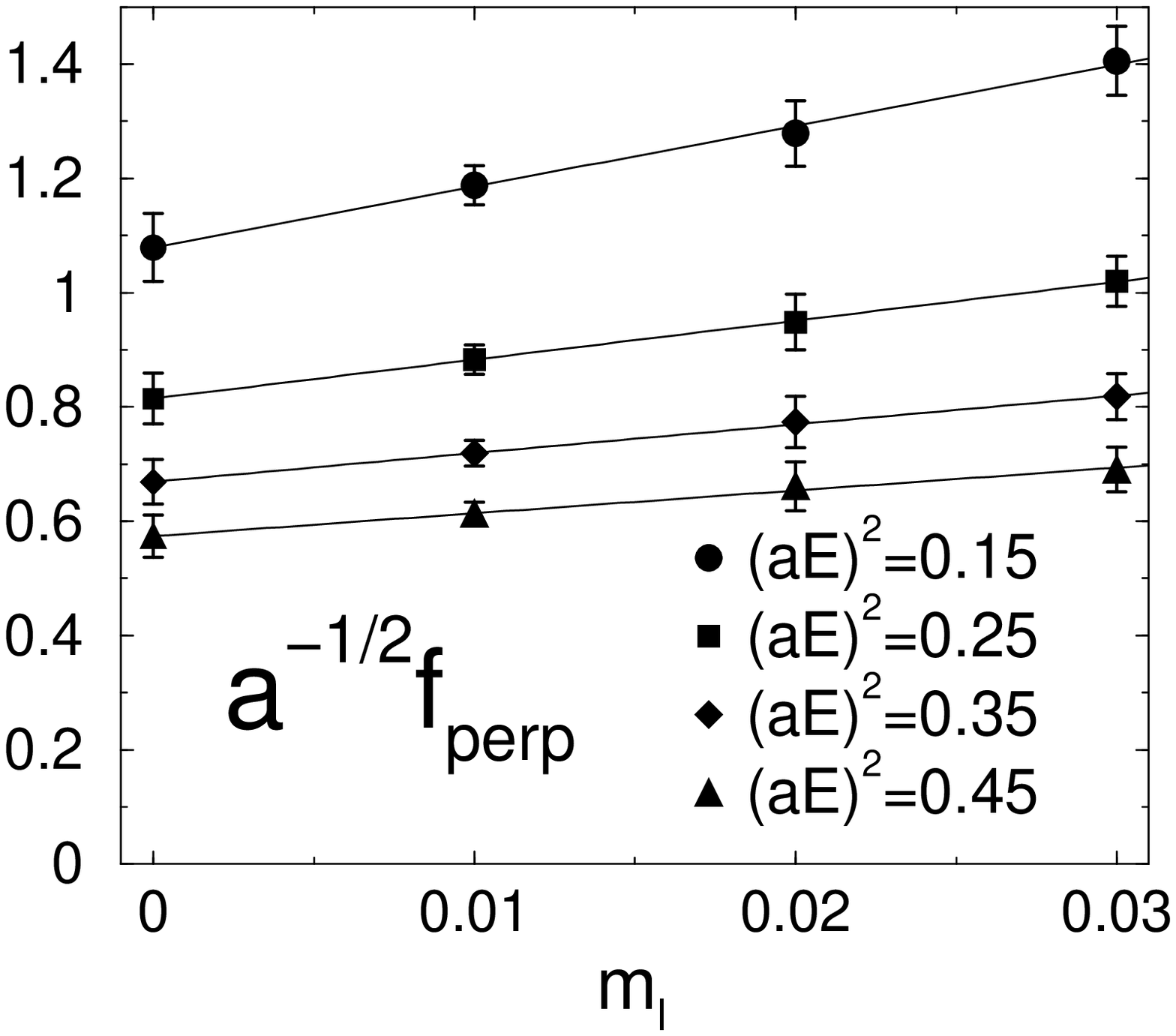}}
\vspace{-1cm}
\caption{Chiral extrapolation for $f_\perp^{D\to\pi}$. }
\vspace{-.8cm}
\label{fig:chiral}
\end{figure}

\begin{figure}[t]
\includegraphics*[width=7.cm]{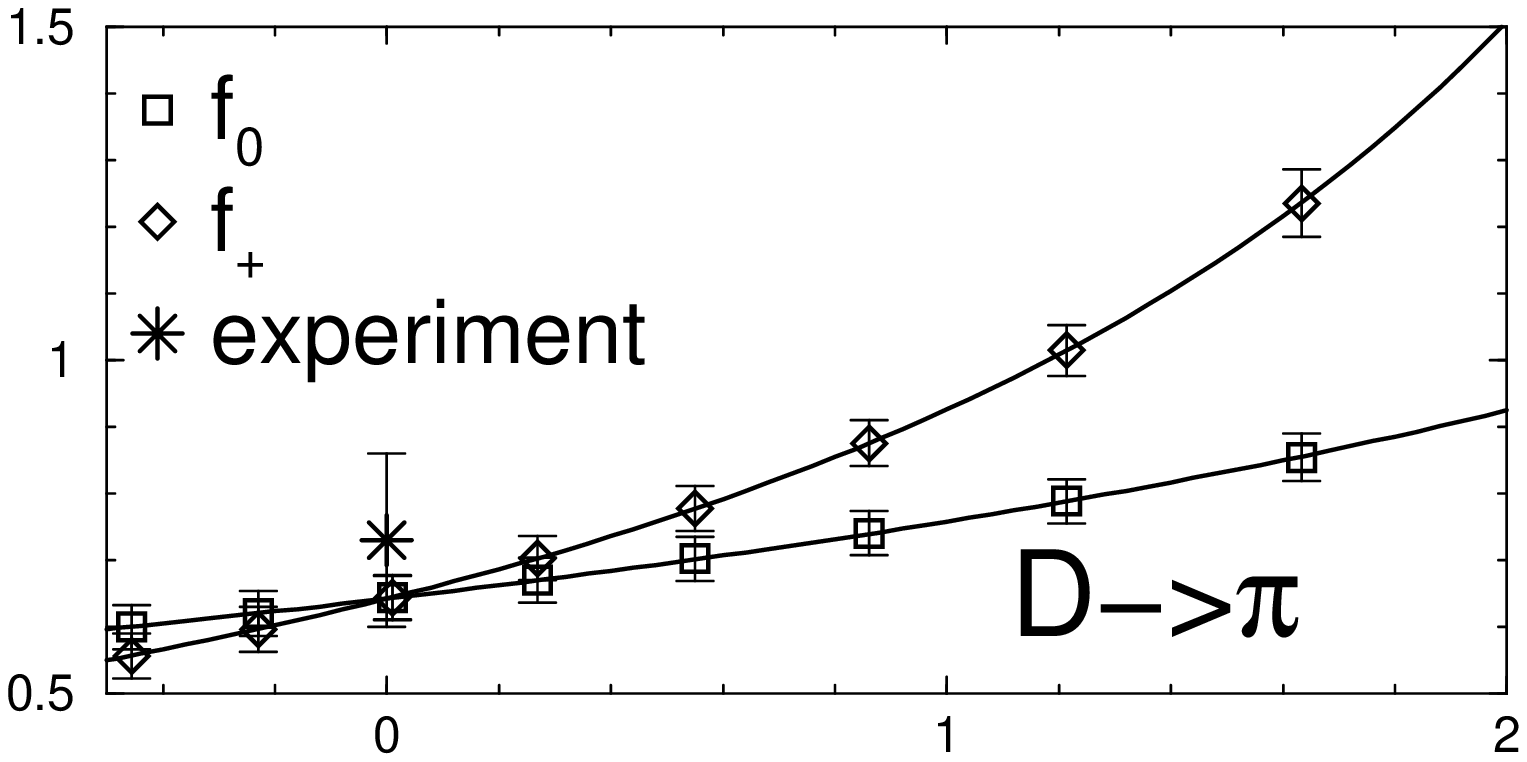}\vspace{-.9cm}
\includegraphics*[width=7.cm]{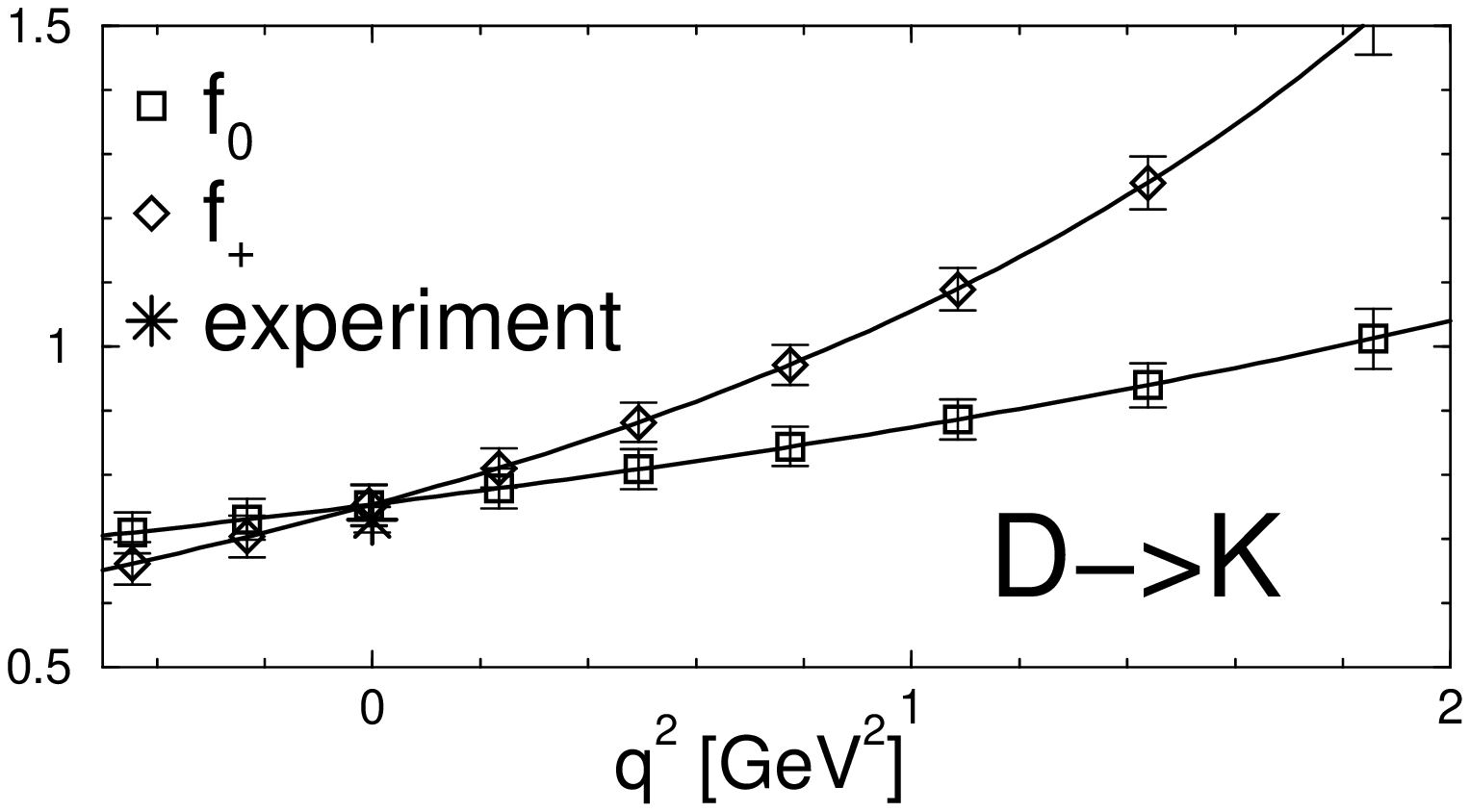}
\vspace{-1.2cm}
\caption{$D\to\pi$ and $D\to K$ form factors.}
\vspace{-.7cm}
\label{fig:D2piK}
\end{figure}

\subsection{$D\to\pi$ and $D\to K$}

To obtain $D\to\pi/K$ form factors at the physical quark mass,
we need to perform a chiral extrapolation
using data in range of $m_l^{\rm val}=0.01$-0.03.
We do this for $f_\parallel$ and $f_\perp$ at fixed pion(kaon) energies
$E_{\pi(K)}$ because the chiral perturbation formulas for the heavy to light
form factors are given in such a way\cite{Becirevic:2002sc}. 
In order to interpolate and extrapolate the results 
to common values of $E_\pi$,
we use a fit with the BK parametrization.
We then perform a linear chiral extrapolation in 
$m_l^{\rm val}$ at nine values of $(aE_\pi)^2$.
One example of these procedures is shown in Fig.~\ref{fig:chiral} for 
$f_\perp^{D\to\pi}$.
Finally $f_\parallel$ and $f_\perp$ 
are converted to $f_0$ and $f_+$.

The $D\to\pi$ and $D\to K$ form factors are shown in 
Fig.~\ref{fig:D2piK} together with experimental values at 
$q^2=0$\cite{Hagiwara:fs}.
Our results at $q^2=0$ are 
\beq
f_+^{D \to K}(0) = 0.75(3) ,~~~ f_+^{D \to \pi}(0) = 0.64(3) 
\eeq
with statistical errors only, whereas experimental values are 
$
f_+^{D \to K}(0) = 0.73(2)
$
and
$
f_+^{D \to \pi}(0) = 0.73(13)
$
with 
$|V_{cs}|=0.996(13)$ and $|V_{cd}|=0.224(16)$\cite{Hagiwara:fs}.
Our results are in agreement with the experimental values.
The analysis including the chiral logarithm and 
the one-loop renormalization constant is underway.

\vspace{.3cm}
\hspace{-.4cm}{\bf Acknowledgments:}
We thank the MILC collaboration for the use of their configurations,
and 
the Fermilab Computing Division and the SciDAC program for their support.
Fermilab is operated by Universities Research Association Inc., under
contract with the DOE.


\end{document}